\documentclass[twocolumn,times]{aastex631}

\usepackage{graphicx}	
\usepackage{amsmath}	

\newcommand{\rgg}{r_\mathrm{gg}}
\newcommand{\pklin}{P_\mathrm{lin}(k)}
\newcommand{\pklinplk}{P_\mathrm{lin}^{\mathrm{CMB}}(k)}
\newcommand{\bg}{b_\mathrm{g}}

\newcommand{\bc}{b_\mathrm{c}}
\newcommand{\rgm}{r_\mathrm{gm}}
\newcommand{\rcg}{r_\mathrm{cg}}
\newcommand{\rcm}{r_\mathrm{cm}}
\newcommand{\ugm}{\Upsilon_\mathrm{gm}}
\newcommand{\ucm}{\Upsilon_\mathrm{cm}}

\newcommand{\ummplk}{\Upsilon_\mathrm{mm}^{\mathrm{CMB}}}
\newcommand{\wpgg}{w_{p,\mathrm{gg}}}

\newcommand{\wpmmplk}{w_{p,{\mathrm{mm}}}^{\mathrm{CMB}}}
\newcommand{\wpcg}{w_{p,\mathrm{cg}}}
\newcommand{\wpcc}{w_{p,\mathrm{cc}}}

\newcommand{\ximm}{\xi_\mathrm{mm}}
\newcommand{\ximmplk}{\xi_\mathrm{mm}^{\mathrm{CMB}}}
\newcommand{\xigg}{\xi_\mathrm{gg}}
\newcommand{\xigm}{\xi_\mathrm{gm}}
\newcommand{\eq}{{=}}

\newcommand{\ds}{\Delta\Sigma}

\newcommand{\dsmmplk}{\Delta\Sigma_\mathrm{mm}^{\mathrm{CMB}}}
\newcommand{\smmplk}{\Sigma_\mathrm{mm}^{\mathrm{CMB}}}

\newcommand{\sig}{\sigma_8}
\newcommand{\sigplk}{\sigma_8^{\rm CMB}}
\newcommand{\Seight}{S_8}

\newcommand{\om}{\Omega_\mathrm{m}}
\newcommand{\omplk}{\Omega_\mathrm{m}^{\rm CMB}}

\newcommand{\hmpc}{h^{-1}\mathrm{Mpc}}

\newcommand{\lgm}{\log M_*}
\newcommand{\ms}{M_*}

\newcommand*\x{{\color{red}\mathrm{X}}}

\begin{document}

\title{Is the large-scale structure traced by the BOSS LOWZ galaxies consistent with {\it Planck}?}

\author[0000-0002-4585-3985]{Zhiwei Shao}
\affiliation{Department of Astronomy, School of Physics and Astronomy, and
Shanghai Key Laboratory for Particle Physics and Cosmology, Shanghai Jiao
Tong University, Shanghai 200240, China; \href{mailto:yingzu@sjtu.edu.cn}{yingzu@sjtu.edu.cn}}

\author[0000-0001-6966-6925]{Ying Zu}
\affiliation{Department of Astronomy, School of Physics and Astronomy, and
Shanghai Key Laboratory for Particle Physics and Cosmology, Shanghai Jiao
Tong University, Shanghai 200240, China; \href{mailto:yingzu@sjtu.edu.cn}{yingzu@sjtu.edu.cn}}
\affiliation{Key Laboratory for Particle Physics, Astrophysics and
Cosmology, Ministry of Education, Shanghai Jiao Tong University, Shanghai
200240, China}

\author[0000-0001-8534-837X]{Huanyuan Shan}
\affiliation{Shanghai Astronomical Observatory (SHAO), Nandan Road 80, Shanghai 200030, China}
\affiliation{University of Chinese Academy of Sciences, Beijing 100049, China}



\begin{abstract}
    Recently, several studies reported a significant discrepancy between
    the clustering and lensing of the Baryon Oscillation Spectroscopic
    Survey~(BOSS) galaxies in the {\it Planck} cosmology. We construct a
    simple yet powerful model based on the linear theory to assess whether
    this discrepancy points toward deviations from {\it Planck}. Focusing
    on scales $10{<}R{<}30\,h^{-1}\mathrm{Mpc}$, we model the amplitudes of
    clustering and lensing of BOSS LOWZ galaxies using three parameters:
    galaxy bias $b_\mathrm{g}$, galaxy-matter cross-correlation coefficient
    $r_\mathrm{gm}$, and $A$, defined as the ratio between the true and
    {\it Planck} values of $\sigma_8$. Using the cross-correlation matrix
    as a diagnostic, we detect systematic uncertainties that drive spurious
    correlations among the low-mass galaxies. After building a clean LOWZ
    sample with $r_\mathrm{gm}{\sim}1$, we derive a joint constraint of
    $b_\mathrm{g}$ and $A$ from clustering+lensing, yielding
    $b_\mathrm{g}{=}2.47_{-0.30}^{+0.36}$ and $A{=}0.81_{-0.09}^{+0.10}$,
    i.e., a $2\sigma$ tension with {\it Planck}. However, due to the strong
    degeneracy between $b_\mathrm{g}$ and $A$, systematic uncertainties in
    $b_\mathrm{g}$ could masquerade as a tension with $A{=}1$. To ascertain
    this possibility, we develop a new method to measure $b_\mathrm{g}$
    from the cluster-galaxy cross-correlation and cluster weak lensing
    using an overlapping cluster sample. By applying the independent bias
    measurement~($b_\mathrm{g}{=}1.76{\pm}0.22$) as a prior, we
    successfully break the degeneracy and derive stringent constraints of
    $b_\mathrm{g}{=}2.02_{-0.15}^{+0.16}$ and $A{=}0.96{\pm}{0.07}$.
    Therefore, our result suggests that the large-scale clustering and
    lensing of LOWZ galaxies are consistent with {\it Planck}, while the
    different bias estimates may be related to some observational
    systematics in the target selection.
\end{abstract}



\section{Introduction}
\label{sec:intro}

Anchored by the latest {\it Planck} observations of the cosmic microwave
background~(CMB) anisotropies at
recombination~\citep{Planck2020parameters}, the standard $\Lambda$CDM
cosmological model under General Relativity~(GR) provides a remarkably good
description of the evolution of our Universe toward later epochs, including
the expansion history measured by baryon acoustic oscillations
\citep{Alam2021} and Type Ia supernovae~\citep{Scolnic2018} as well as the
growth history measured by redshift space distortion, cosmic shear, and
galaxy clusters~\citep[see][for an extensive review]{Weinberg2013}.
However, tensions may still arise when one compares low-redshift
measurements of the matter density $\om$ and the present-day amplitude of
matter clustering, characterized by $\sig$, the rms matter fluctuation in
$8\,\hmpc$ spheres, to the values expected from extrapolating CMB
anisotropies forward from recombination to $z{=}0$, e.g., $\omplk{=}0.3153$
and $\sigplk{=}0.8111$ using {\it Planck}.

Most notably, recent cosmic shear studies found a $1{-}3\,\sigma$ lower
value~\citep{Asgari2021KiDS,Secco2022DES,Amon2022DES, Huterer2022} of the
parameter combination $\Seight{\equiv}\sig(\om/0.3)^{0.5}$ compared to {\it
Planck}~\citep[but see][for a plausible non-linear
solution]{Amon2022lows8nonlinear}. An alternative manifestation of this
$\Seight$-tension is the apparent mismatch between the clustering and
galaxy-galaxy~(g-g) lensing of the Baryon Oscillation Spectroscopic
Survey~(BOSS) galaxies when assuming {\it Planck} cosmology~\citep[a.k.a.,
lensing-is-low;][]{Leauthaud2017, Lange2019}. In this letter, we examine
the consistency~(or lack thereof) between $\sigplk$ and the $\sig$ measured
from the clustering and g-g lensing of BOSS LOWZ galaxies over scales
between $10$ and $30\,\hmpc$. In particular, we elucidate the role of the
galaxy bias $\bg$ in this consistency test by introducing an independent
prior on $\bg$, measured with the help of an overlapping sample of
clusters.

Although the existence of a lensing-is-low effect on scales below
${\sim}5\,\hmpc$ remains a subject of intense debate~\citep{More2015,
Yuan2020MNRASAssmeblyBias, Lange2021, ChavesMontero2022,
Contreras2022consistentGCGGL}, using data from three different lensing
surveys~\citet{Amon2023} demonstrated that on scales above $5.25\,\hmpc$,
there exists a $2.3\,\sigma$ discrepancy between clustering and g-g lensing
in the {\it Planck} cosmology for the entire BOSS
sample~(${\sim}1.5\,\sigma$ for the LOWZ galaxies between
$0.15{<}z{<}0.31$). Compared with the lensing-is-low effect on small
scales, the large-scale discrepancy is a more ``direct'' tension with {\it
Planck} --- it does not depend on the complex modelling of galaxy-halo
connection, which on small scales is plagued by galaxy assembly bias and
baryonic feedback~\citep[][]{Salcedo2022, Beltz-Mohrmann2022}.  Therefore,
it is imperative that this direct tension be assessed in a simple framework
that confronts the LOWZ clustering+lensing measurements with the prediction
by the {\it Planck} $\Lambda$CDM+GR model at asymptotically large
scales~(i.e., ${>}10\,\hmpc$), where structure growth follows the linear
theory and galaxy bias becomes scale-independent.

On scales above $10\,\hmpc$, galaxy clustering and g-g lensing are
measuring
\begin{equation}
    \xigg=\bg^2\ximm\propto\bg^2\sig^2
    \label{eqn:xigg}
\end{equation}
    and
\begin{equation}
    \xigm=\bg\rgm\ximm\propto\bg\rgm\sig^2,
    \label{eqn:xigm}
\end{equation}
respectively, where $\ximm$ is the matter correlation and
$\rgm{\rightarrow}1$ is the cross-correlation coefficient between galaxies
and matter~\citep{Cacciato2012}. Note that we fix the value of $\om$ to be
$\omplk$, so that the $\Seight$-tension
simplifies into a $\sig$-tension. A joint
analysis of clustering and g-g lensing can then measure
\begin{equation}
\xigm/\sqrt{\xigg}=\rgm\sig\rightarrow\sig,
    \label{eqn:cancel}
\end{equation}
thereby cancelling the unknown\footnote{The cancellation, however, requires
$\bg{=}\xigm/\xigg$, which puts a constraint on the galaxy-halo
connection if the small scales are included.} nuisance parameter $\bg$.
Thus, a strong discrepancy between clustering and g-g lensing in {\it
Planck} is usually interpreted as the evidence of the ratio
\begin{equation}
A \equiv \sig/\sigplk,
    \label{eqn:A}
\end{equation}
significantly deviating from unity~(e.g., $A{<}1$ means lensing-is-low).
For instance, \citet{Wibking2020} constructed a Halo Occupation
Distribution-based nonlinear emulator to constrain cosmology from jointly
modelling the clustering and g-g lensing of the LOWZ galaxies on scales
above $0.6\,\hmpc$, finding a $3.5\,\sigma$~($2.6\,\sigma$ if limited to
$>2\,\hmpc$) evidence of $A{<}1$.
By explicitly modelling g-g lensing in the form of
$\om\rgm\sqrt{\ximm\xigg}$, \citet{Singh2020} found a similar
${\sim}3\,\sigma$ discrepancy with {\it Planck} for a minimum scale of
$2\,\hmpc$ using the LOWZ galaxies. They also found the discrepancy
persists at the level of ${\sim}1.5\,\sigma$ when limited to scales
$>10\,\hmpc$.

Alternatively, a clustering-lensing mismatch could be the result of having
systematic errors that drive $\rgm$ below unity or/and an incorrect
clustering amplitude, leading to an imperfect cancellation of $\bg$ in
Equation~\ref{eqn:cancel}. Although both possibilities are generally
considered unlikely for well-defined galaxy samples and are thus omitted in
previous studies, \citet{Zu2020} found that the clustering of LOWZ galaxies
on scales above $10\,\hmpc$ exhibits a non-monotonic trend with stellar
mass $\ms$, with the low-$\ms$ bin having a higher clustering amplitude
than the intermediate-$\ms$ one.  \citet{Zu2020} interpreted this
clustering anomaly as the evidence for the low-$\ms$ galaxies being the
satellites of massive haloes. In this letter, we demonstrate that the
clustering anomaly is instead caused by some unknown systematic
uncertainties associated with the BOSS LOWZ sample, probably due to the
complex target selection criteria of BOSS galaxies~\citep{Reid2016}.

After building a clean sample of LOWZ galaxies free of the clustering
anomaly in \S\ref{sec:cleanlowz}, we reproduce the lensing-is-low effect
using our linear framework in \S\ref{sec:simple}.  After developing a novel
method of measuring $\bg$ from an overlapping sample of galaxy clusters, we
demonstrate in \S\ref{sec:cluster} that by applying this measurement as a
prior on $\bg$, we can mitigate the impact of systematic uncertainties in
$\bg$ and resolve the direct tension between $\sig$ and $\sigplk$
originated from the large scales. We summarize our results and look to the
future in \S\ref{sec:conc}.

\section{A ``Clean'' Sample of LOWZ Galaxies}
\label{sec:cleanlowz}

\subsection{The BOSS LOWZ galaxies}
\label{subsec:lowz}

As part of the SDSS-III programme~\citep{Eisenstein2011},
BOSS~\citep{Dawson2013} observed the spectra of 1.5 million galaxies over a
sky area of $\sim$10000 deg$^2$ at $0.15{<}z{<}0.7$.  The BOSS targets were
selected from the Data Release 8~\citep[DR8;][]{Aihara2011} of SDSS
five-band imaging, using two separate sets of colour and magnitude cuts for
the LOWZ~($0.15{<}z{<}0.43$) and CMASS~($0.43{<}z{<}0.7$)
samples~\citep{Reid2016}. We use the DR12 of the BOSS
LOWZ sample~\citep{Alam2015} and limit our analysis to the Northern Galactic
Cap. We adopt the aperture-corrected stellar mass measurements by
\citet{Chen2012StellarMass}, but reduce the stellar mass values by 0.155
dex to be consistent with the SDSS main galaxies at
$z{<}0.1$~\citep{Guo2018CSMF}.

In this letter, we focus on the LOWZ galaxies in the redshift range
$z{=}[0.2, 0.3]$, for which we have a
volume-complete sample of photometric clusters with excellent photo-z
accuracy in the same footprint from redMaPPer~\citep{Rykoff2014redmapper}.
We use 4580 clusters with richness $\lambda$ above $20$ from the SDSS
redMaPPer v6.3 catalogue~\citep{Rykoff2016redmapper}. Our results do not
change when using a higher threshold of $\lambda{=}30$ or only clusters
with spectroscopic redshifts.  As will be demonstrated later in
\S\ref{sec:cluster}, this overlapping cluster sample allows us the unique
opportunity to make an independent measurement of $\bg$, without resorting
to galaxy clustering or g-g lensing.

\subsection{Selecting a clean sample of LOWZ galaxies}
\label{subsec:clean}

\begin{figure}
    \centering
    \includegraphics[width=0.96\linewidth]{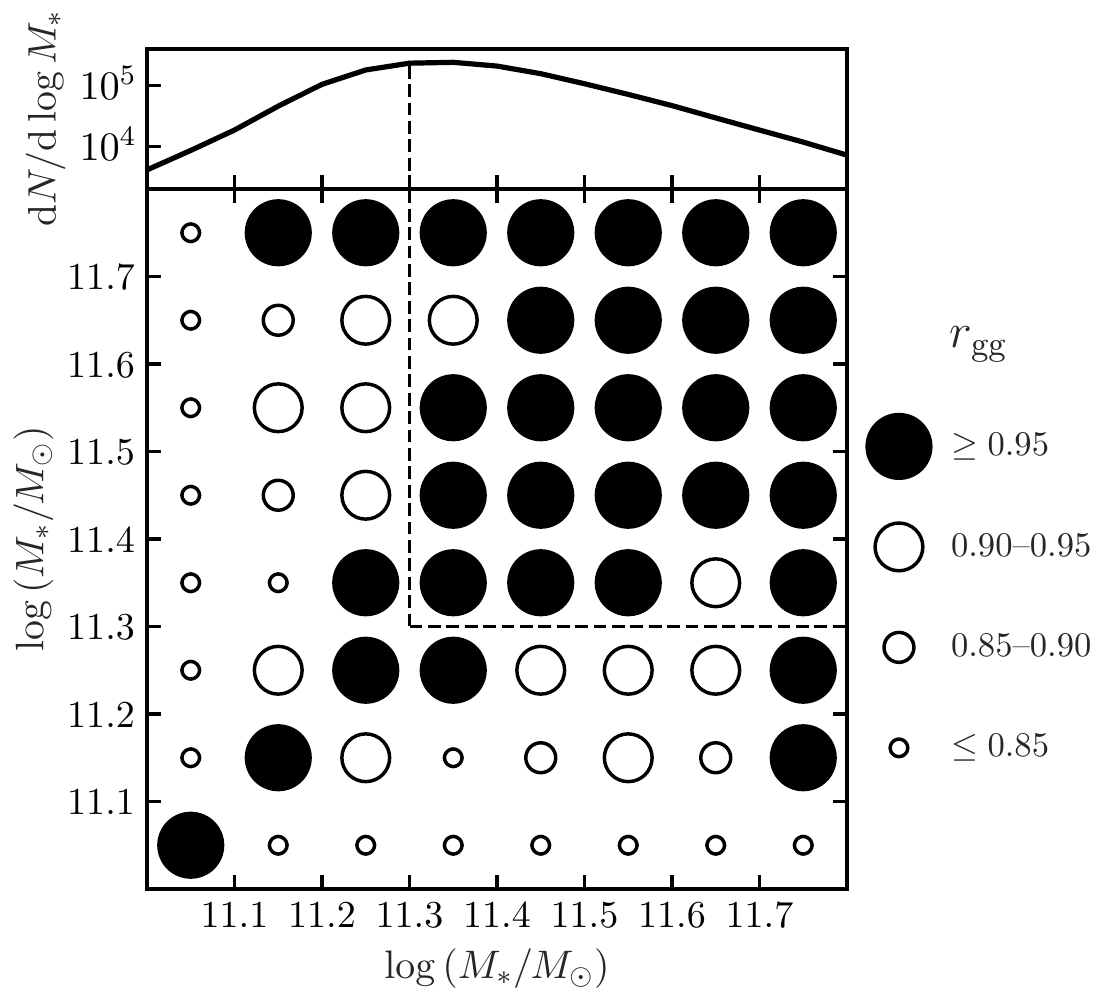}
    \caption{Cross-correlation matrix $\mathcal{Q}$. The size of each
    circle corresponds to the value of the cross-correlation coefficient
    $\rgg$ between two stellar mass bins, indicated by the four circles on
    the right. Filled circles represent near-perfect correlations with
    $\rgg{\geq}0.95$. The black solid curve on top shows the stellar mass
    distribution of the LOWZ galaxies, with the black dashed
    line indicating the stellar mass limit~($\lgm{=}11.3$) of our C-LOWZ
    sample selection.} \label{fig:rgg}
\end{figure}

Ideally, we hope to select a ``clean'' LOWZ sample that has $\rgm\,{=}\,1$
on scales above $10\,\hmpc$, but $\rgm$ is not directly accessible.
Following an indirect diagnostic proposed by \citet{Tegmark1999}, we divide
the overall LOWZ galaxy sample into eight narrow bins by stellar
mass~($\Delta\log\ms{=}0.1\,\mathrm{dex}$), and then compute the
galaxy-galaxy cross-correlation coefficient $\rgg^{ij}$ between each pair
combination $(i,j)$, yielding an $8{\times}8$ cross-correlation matrix
$\mathcal{Q}\equiv[\rgg^{ij}]$. We search $\mathcal{Q}$ for a contiguous
block within which the average $\rgg$ is close to unity, and identify the
stellar mass range of that block as a {\it clean} sample. The rationale is
as follows. If the $i$-th and $j$-th bins are both perfectly correlated
with matter, then they must also be perfectly correlated with each
other~(i.e., $\rgg^{ij}{=}1$).  Conversely, if we find imperfect
correlations between the two, then $\rgm{<}1$ must be true for at least one
of the two bins. Therefore, by culling the $\rgg{<}1$ regions from
$\mathcal{Q}$, we remove the $\rgm{<}1$ galaxies that are potentially
subjected to observational systematics.

We compute $\rgg^{ij}$ as
\begin{equation}
    \rgg^{ij} = \frac{w_p^{ij}}{\sqrt{w_p^{ii} \,w_p^{jj}}}\,\Bigg|_{10 < R < 30},
    \label{eqn:rgg}
\end{equation}
where $w_p^{ii}$ and $w_p^{jj}$ are the projected auto-correlation
functions of the $i$-th and $j$-th bins, respectively, and $w_p^{ij}$ is
the projected cross-correlation function between the two bins, all
evaluated over the projected separation of $R{\in}[10,\,30]\,\hmpc$.  We
compute the projected correlation functions $w_p(R)$ by integrating the 2D
redshift-space correlation function $\xi^{rs}(R, \Pi)$ along the
line-of-sight distance $\Pi$
\begin{equation}
    w_{p}(R) = \int_{-\Pi_\mathrm{max}}^{+\Pi_\mathrm{max}}\xi^{rs}(R, \Pi)\operatorname{d}\Pi,
    \label{eqn:wpobs}
\end{equation}
where we set the integration limit $\Pi_\mathrm{max}$ to $50\,\hmpc$.  For
measuring $\xi^{rs}(R, \Pi)$, we use the Landy-Szalay
estimator~\citep{Landy1993LSEstimator} and estimate the associated
uncertainty matrix by applying the jackknife re-sampling method over 128
sub-divided regions across the footprint.

Figure~\ref{fig:rgg} visualizes the correlation matrix $\mathcal{Q}$, with
the size of each circle at column $i$ and row $j$ representing the value of
$\rgg^{ij}$, which we classify into four categories listed on the right.
Surprisingly, almost all of the matrix elements that involve bins with
$\log\ms{<}11.3$ are below $0.95$, hinting at the existence of unknown
systematics among the low-$\ms$ galaxies. The highest-$\rgg$
category~($\rgg{>}0.95$) is highlighted by the filled circles, which are
mostly enclosed within the dashed lines, i.e., $\log\ms{\geq}11.3$.  The
dashed line also coincides with the peak of the stellar mass distribution
of the LOWZ sample~(top), suggesting that the unknown
systematics are likely from the target selection in the low-mass
range with high incompleteness. In particular, we detect significant spurious
correlations over large line-of-sight distances within the galaxies
with $\log\ms{<}11.3$, which explains the aforementioned clustering anomaly
found in \citet{Zu2020}.

Informed by the $\mathcal{Q}$ diagnostics, we select LOWZ galaxies with
$\log\ms{\geq}11.3$ into our ``clean'' LOWZ
sample~(hereafter referred to as C-LOWZ), and use the C-LOWZ sample for all
subsequent analyses in this letter.

\subsection{Clustering and weak lensing measurements}
\label{subsec:measurements}

\begin{figure*}
    \centering \includegraphics[width=.96\linewidth]{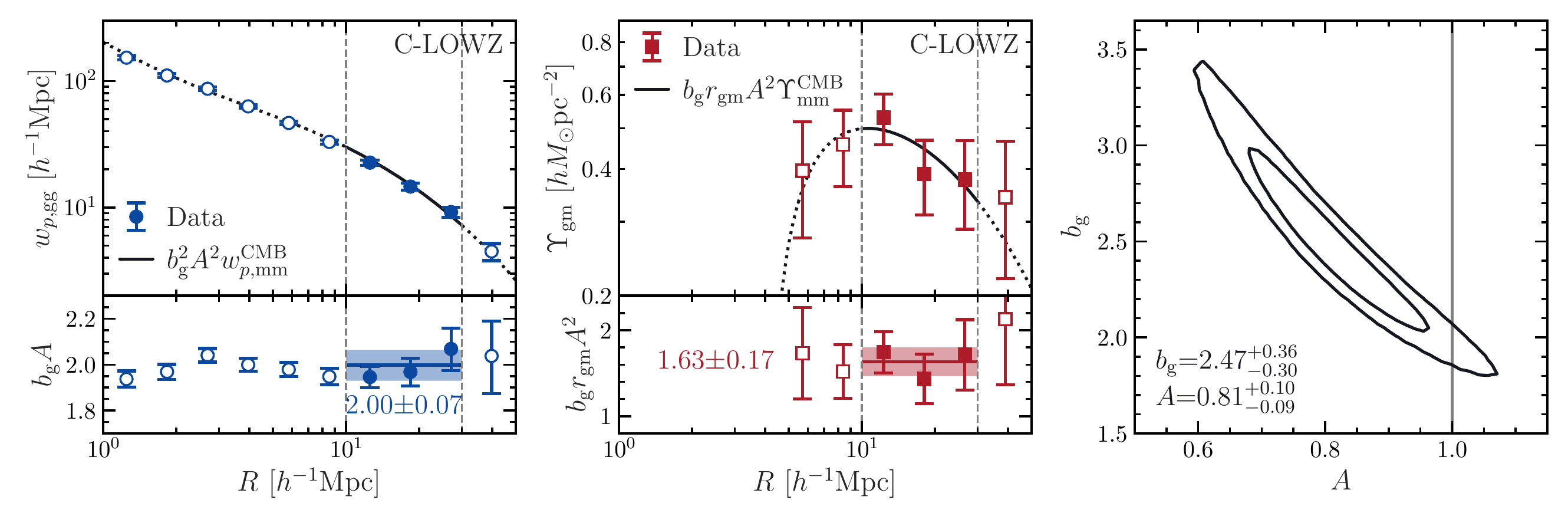}
    \caption{{\it Left}: Comparison of the projected clustering of C-LOWZ
    galaxies between the measurements~(circles with errorbars) and the
    best-fitting linear model prediction~(solid curve) on scales
    $10{<}R{<}30\,\hmpc$~(within the vertical dashed lines). The
    constraint on the product of $\bg A$ is shown by the horizontal band in
    the bottom panel.  {\it Middle}: Similar to the left panel, but for the
    g-g lensing signal $\ugm$.  {\it Right}: Joint
    posterior distribution of $\bg$ and $A$ derived from combining the clustering and
    g-g lensing of C-LOWZ galaxies. Contours indicate the 68\% and 95\%
    2D confidence regions, with the marginalized $1\sigma$ constraints listed in
    the bottom left.} \label{fig:gg_wp_upsilon}
\end{figure*}

We make use of the projected auto-correlations of C-LOWZ galaxies~($\wpgg$)
and redMaPPer clusters~($\wpcc$), as well as the cross-correlation between
the two~($\wpcg$) in our analyses. The three correlation functions $w_p(R)$
are measured in the same way as described by Equation~\ref{eqn:wpobs}. For
the weak lensing by C-LOWZ galaxies and redMaPPer clusters, we measure the
surface density contrast profile $\ds(R)$ using the shear catalogue derived
from the Dark Energy Camera Legacy Survey~\citep[DECaLS;][]{Dey2019}.  The
same DECaLS data were used in the fiducial cluster weak lensing study of
\citet{Zu2021}, who found excellent agreement between the $\ds$ measured
from DECaLS and that from SDSS imaging using the re-Gaussianization
algorithm \citep{Reyes2012}, but the uncertainties in the DECaLS
measurements are smaller by roughly a factor of two thanks to the
deeper depth. Note that the SDSS g-g lensing measurements were used by
\citet{Wibking2020}, \citet{Singh2020}, and \citet{Lange2021} in their LOWZ
analyses.

Unlike $w_p(R)$, the $\ds(R)$ signal on scales above $10\,\hmpc$ is
contaminated by the non-linear structure growth below
$10\,\hmpc$~\citep{Zu2015}. To remedy this, we measure the annular
differential surface density~(ADSD) $\Upsilon$, defined as
\begin{equation}
        \Upsilon(R)=\ds(R)-\left(\frac{R_0}{R}\right)^2\ds(R_0),
        \label{eqn:ups}
\end{equation}
where we set the minimum scale $R_0{=}4\,\hmpc$. Originally proposed by
\citet{Baldauf2010Upsilon}, $\Upsilon$ removes all the information from
scales below $R_0$ and thus can be safely predicted using the linear theory.
Following \citet{Mandelbaum2013}, we interpolate the value of $\ds(R_0)$
using the best-fitting power-law that describes $\ds$ between $1\,\hmpc$
and $10\,\hmpc$~\citep[also see][]{Singh2020}.

\section{Linear Assessment Method}
\label{sec:simple}

To facilitate the $\sig$-tension assessment on linear scales, we follow the
practice of \citet{Zu2014} and keep the shape of the linear matter power
spectrum $\pklin$ fixed to that of {\it Planck} $\pklinplk$.
With a fixed $\pklin$ shape, a single value of $A$ specifies the full
matter correlation function $\ximm$ via
\begin{equation}
    \ximm(r) = A^2 \ximmplk(r).
\end{equation}
where $\ximmplk(r)$ is the nonlinear matter correlation predicted from
$\pklinplk$ using the prescription of~\citet{Takahashi2012}. We assume
the mean redshift of C-LOWZ~($z{=}0.256$) for all calculations.

Focusing exclusively on scales $10{<}R{<}30\,\hmpc$, we model the
$\wpgg$ measurement as
\begin{equation}
    \wpgg(R) =  \bg^2 \, A^2 \, f_\mathrm{rrsd}(R\mid\bg) \; \wpmmplk(R),
    \label{eqn:wpgg}
\end{equation}
where $\wpmmplk(R)$ is the projected matter auto-correlation
\begin{equation}
    \wpmmplk(R) = 2\int_{R}^{\sqrt{\Pi_\mathrm{max}^2
    + R^2}}\ximmplk(r)\frac{r\operatorname{d}r}{\sqrt{r^2-R^2}},
\end{equation}
and $f_\mathrm{rrsd}(R|\bg)$ accounts for the
scale-dependent enhancement of $\wpgg(R)$ due to the residual
redshift-space distortion~(RRSD) effect.
We calculate $f_\mathrm{rrsd}$ using the modified linear Kaiser formalism of
\citet{vandenBosch2013rrsd}. For the C-LOWZ sample, $f_\mathrm{rrsd}$ is
about 8\% and 30\% at $R{=}10$ and $30$ $\hmpc$, respectively.
The $\bg$-dependence of $f_\mathrm{rrsd}$ is very weak and does not provide
any meaningful constraint on $\bg$.  Therefore, our clustering model of
Equation~\ref{eqn:wpgg} constrains the parameter combination $\bg A$, which
turns into a bias measurement $\bg^{\mathrm{clustering}}$ for any given
value of $A$.

\begin{figure*}
    \centering \includegraphics[width=.96\linewidth]{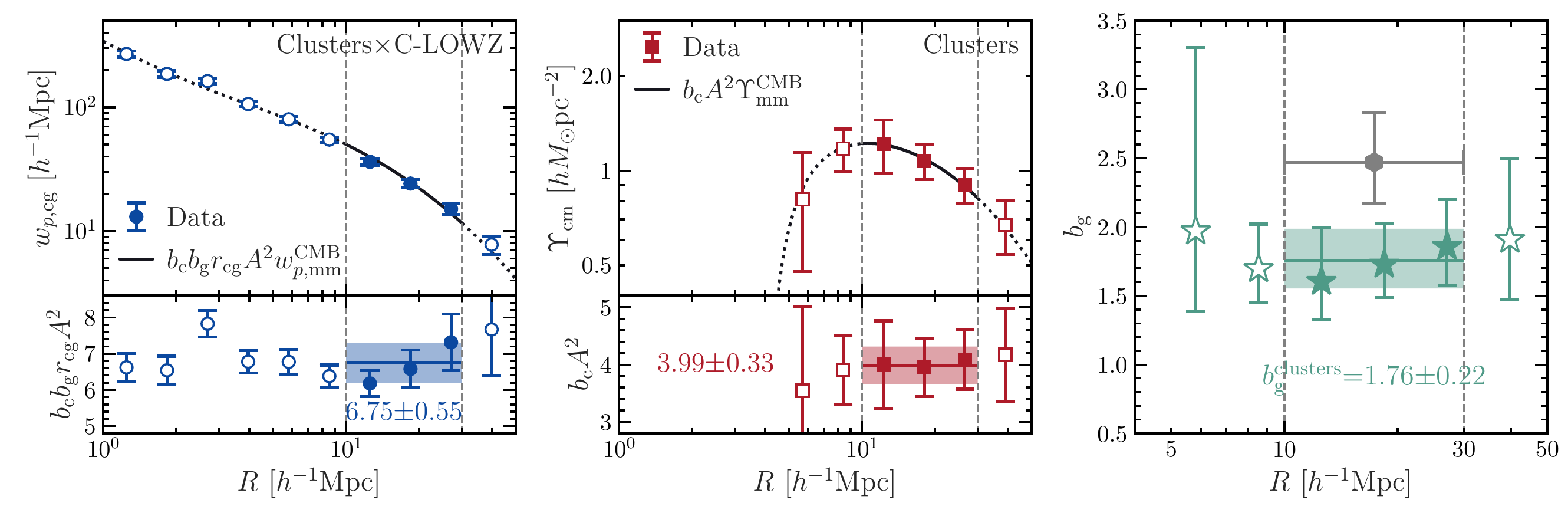}
    \caption{Cluster-based bias measurement for the C-LOWZ galaxies~(right
    panel) from the combination of cluster-galaxy cross-correlation~(left)
    and cluster weak lensing~(middle).  The formats of the left and middle
    panels are the same as in Figure~\ref{fig:gg_wp_upsilon}.  In the right
    panel, star symbols with errorbars represent the ratio calculated using
    Equation~\ref{eqn:cgratio}, from
    which we derive $\bg^{\mathrm{clusters}}\eq1.76{\pm}0.22$~(green
    horizontal band). Gray hexagon with errorbar indicates the bias
    constraint~($2.47_{-0.30}^{+0.36}$) from the right panel of
    Figure~\ref{fig:gg_wp_upsilon}.} \label{fig:cg_wp_upsilon}
\end{figure*}

By the same token, the ADSD profile $\ugm$ can be modelled as
\begin{equation}
    \ugm(R) = \bg \,\rgm\, A^2\, \ummplk(R),
    \label{eqn:ugm}
\end{equation}
and the matter-matter ADSD in {\it Planck} $\ummplk$ can be predicted from
Equation~\ref{eqn:ups} using $\dsmmplk$
\begin{equation}
    \dsmmplk(R)=\langle\smmplk(<R)\rangle-\smmplk(R),
\end{equation}
where
\begin{equation}
    \smmplk(R)=\omplk{\rho}_\mathrm{crit}
    \int_{-\infty}^{+\infty}\left[1+\ximmplk(r)\right]\frac{r\operatorname{d}r}{\sqrt{r^2-R^2}},
\end{equation}
and
\begin{equation}
\langle\smmplk(<R)\rangle=\frac{2}{R^2}\int_0^R\smmplk(R')R'\operatorname{d}R'.
\end{equation}
Similarly, our lensing model of Equation~\ref{eqn:ugm} constrains the
parameter combination $\bg\rgm A^2$, which then provides a bias measurement
$\bg^{\mathrm{lensing}}$ for any given $\rgm$ at fixed $A$.

Combining the clustering model~(Equation~\ref{eqn:wpgg}) and the lensing
model~(Equation~\ref{eqn:ugm}), our linear assessment model has only three
parameters: $\bg$, $\rgm$, and $A$. Among the three, $A$ is our key
parameter which quantifies the $\sig$-tension, $\bg$ is our primary
nuisance parameter that may still be affected by observational systematics,
and $\rgm$ should be close to unity for our C-LOWZ galaxy sample.  We
perform a suite of three Bayesian inference analyses by applying the
clustering model, the lensing model, and the combined
model to the measurements of $\wpgg$, $\ugm$, and $\wpgg{+}\ugm$,
respectively, of the C-LOWZ sample on scales $10{<}R{<}30\,\hmpc$. Assuming
Gaussian likelihoods, we can derive the posterior distributions of $\bg A$,
$\bg \rgm A^2$, and the joint distribution of $\bg$ and $A$. For the joint
constraint, we apply a conservative Gaussian prior on
$\rgm{\sim}\mathcal{N}(1,0.05^2)$.

Figure~\ref{fig:gg_wp_upsilon} presents the results of the
clustering-only~(left), lensing-only~(middle), and joint~(right) constraints.
On the left panel, the $\wpgg$ measurement~(circles with errorbars) is well
described by the prediction from the best-fitting clustering model~(solid
curve) on scales $10{<}R{<}30\,\hmpc$. Dotted curve segments
are the extrapolation of the linear model into the nonlinear
or very low signal-to-noise regimes, and should not
be compared with the data~(open circles).
The bottom subpanel shows the
square root of the ratio between the measurement and $\wpmmplk$, which
gives the constraint on $\bg A$ as $2.00{\pm}0.07$~(horizontal band). Similarly, the middle panel presents the data vs.
prediction comparison in the main panel, as well as the ratio profile in
the bottom subpanel, yielding a constraint of $\bg\rgm
A^2\eq1.63{\pm}0.17$. If assuming the {\it Planck} cosmology~(i.e.,
$A\eq1$), we would obtain $\bg^{\mathrm{clustering}}\eq2.00{\pm}0.07$ and
$\bg^{\mathrm{lensing}}\eq1.63{\pm}0.17$, reproducing a
lensing-is-low effect of $2\sigma$ on large scales.

The right panel of Figure~\ref{fig:gg_wp_upsilon} shows the 2D
constraint on the $A$ vs. $\bg$ plane after marginalizing over the
uncertainties of $\rgm$. The contour lines
are the 68\% and 95\% confidence limits, yielding the 1D posterior constraints
as $\bg\eq2.47_{-0.30}^{+0.36}$ and $A\eq0.81_{-0.09}^{+0.10}$.  Therefore,
when applied to the clustering and g-g lensing measurements on large
scales, our linear assessment method detects a ${\sim}2\sigma$ discrepancy
with {\it Planck}, reproducing the $\Seight$-tension reported by
previous studies.

However, a strong degeneracy exists between $\bg$ and $A$, and the 2D
constraint remains marginally consistent with {\it Planck} within
$1.5\sigma$.  Given the presence of unknown systematics in the LOWZ sample
demonstrated by Figure~\ref{fig:rgg}, it is plausible that other
systematics of the similar origin may remain within our C-LOWZ sample in a
way that affects the $\bg$ measurement but evades our detection using the
\citet{Tegmark1999} method.  In order to break the strong degeneracy seen
in the right panel of Figure~\ref{fig:gg_wp_upsilon}, we need to place an
external prior on $\bg$ using an independent measurement of $\bg$ from a
different dataset than galaxy clustering and g-g lensing.

\section{Cluster-based Measurement of $\bg$}
\label{sec:cluster}

Adopting the same philosophy of \S\ref{sec:simple}, we can model the
projected cluster-galaxy cross-correlation $\wpcg$ as
\begin{equation}
    \wpcg(R) =  \bc\,\bg\,\rcg\,A^2 \,f_\mathrm{rrsd}(R\mid\bc,\bg) \; \wpmmplk(R),
    \label{eqn:wpcg}
\end{equation}
where $\bc$ is the cluster bias, $f_\mathrm{rrsd}$ accounts for the
cluster-galaxy RRSD, and $\rcg$ is the cluster-galaxy
cross-correlation coefficient. We can directly measure $\rcg$ via
\begin{equation}
    \rcg = \frac{\wpcg}{\sqrt{\wpcc\,\wpgg}}\,\Bigg|_{10 < R < 30},
    \label{eqn:rcg}
\end{equation}
where $\wpcc$ is the projected auto-correlation of clusters. Likewise, the cluster weak lensing $\ucm$ can be modelled as
\begin{equation}
    \ucm(R) = \bc \,\rcm\, A^2\, \ummplk(R),
    \label{eqn:ucm}
\end{equation}
where $\rcm$ is the cluster-matter cross-correlation coefficient. For
the massive clusters that are robustly detected from imaging~(i.e., without
spectroscopic target selection), we can safely assume $\rcm$ to be unity.
Therefore, by combining the two measurements we can cancel $\bc$ to obtain
\begin{equation}
    \bg^{\mathrm{clusters}} =
    \frac{1}{f_\mathrm{rrsd}\, \rcg}
    \left(\frac{\wpcg}{\wpmmplk}\right)
    \left(\frac{\ucm}{\ummplk}\right)^{-1},
    \label{eqn:cgratio}
\end{equation}
i.e., a cluster-based $\bg$ measurement that is independent of $A$, galaxy
clustering, and g-g lensing\footnote{Although the g-g and cluster lensing are
measured from the same shear catalogue, the uncertainties are both
dominated by the shape noise, so that the two lensing signals can be
considered independent.}.

We apply the new bias measurement method to the C-LOWZ galaxies and the
overlapping redMaPPer clusters, with the results shown in
Figure~\ref{fig:cg_wp_upsilon}.  The left and middle panels illustrate the
constraints from the clustering-only and lensing-only measurements,
respectively, with the same format as that of
Figure~\ref{fig:gg_wp_upsilon}.  In particular, the clustering constraint
on the parameter combination $\bc\bg\rcg A^2$ is $6.75{\pm}0.55$, and the
lensing constraint is $\bc A^2 {=}3.99{\pm}0.33$.  With
$\rcg\eq0.96{\pm}0.03$ measured via Equation~\ref{eqn:rcg}, we can compute
$\bg$ via Equation~\ref{eqn:cgratio}.  Our new independent constraint on
$\bg$ is $\bg^{\mathrm{clusters}}\eq1.76{\pm}0.22$, indicated by the
horizontal green band in the right panel of Figure~\ref{fig:cg_wp_upsilon}.
Compared to our previous constraint using galaxy clustering and g-g
lensing~(gray hexagon with errorbar), the new $\bg^{\mathrm{clusters}}$
measurement is significantly lower, signalling a $2\sigma$ tension in
$\bg$.

\begin{figure}
    \centering
    \includegraphics[width=0.96\linewidth]{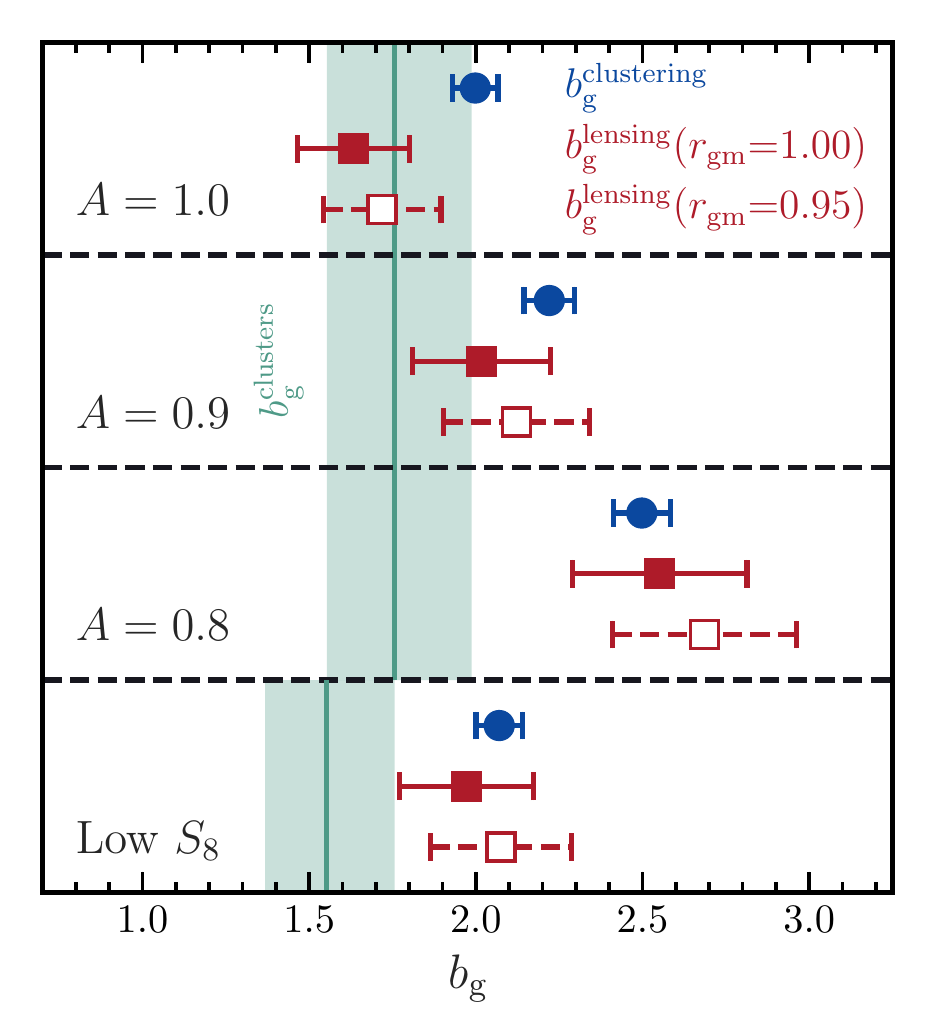}
    \caption{Comparison between galaxy biases measured from the
    cluster-based method~(green vertical band), galaxy clustering~(blue
    circles), and g-g lensing~(red squares), under four different
    cosmologies, including {\it Planck}~($A{=}1.0$), two low-$\sig$
    cosmologies with $A{=}0.9$ and $0.8$, and a representative
    low-$\Seight$ cosmology that resolves the clustering-lensing mismatch.
    Filled~(open) red squares are the $\bg^{\mathrm{lensing}}$ estimates when
    assuming $\rgm{=}1$~($0.95$).} \label{fig:biases}
\end{figure}

Figure~\ref{fig:biases} compares the $\bg^{\mathrm{clusters}}$ measurement,
which is $A$-independent~(green band), to the clustering-only
$\bg^{\mathrm{clustering}}$~(blue circles) and
lensing-only~$\bg^{\mathrm{lensing}}$~(red squares) measurements assuming
four different cosmologies. For the lensing-only measurements, we assume
$\rgm\eq 1$ and $0.95$ for the filled and open squares, respectively. In
the case of {\it Planck}~($A\eq 1$), although $\bg^{\mathrm{clustering}}$
and $\bg^{\mathrm{lensing}}$ are mutually discrepant beyond $1\sigma$, both
are individually consistent with $\bg^{\mathrm{clusters}}$ well within
$1\sigma$. As we move away from {\it Planck} by decreasing $A$, the
$\bg^{\mathrm{clustering}}$ and $\bg^{\mathrm{lensing}}$ values increase by
different amounts so as to become more consistent with each other, echoing
the claim that a low-$\sig$ cosmology could resolve the lensing-is-low
tension; Meanwhile, however, both bias values become progressively more
discrepant with $\bg^{\mathrm{clusters}}$, which does not vary with $A$.
Therefore, although a low-$\sig$ cosmology could resolve the discrepancy
between $\bg^{\mathrm{clustering}}$ and $\bg^{\mathrm{lensing}}$, it
exacerbates the tension of the two biases with $\bg^{\mathrm{clusters}}$.

The strong discrepancy between $\bg^{\mathrm{clusters}}$ and the other two
biases persists when we adopt a representative low-$\Seight$
cosmology~(bottom panel) that alleviates the lensing-is-low problem, with
the parameters drawn from the table 2 of \citet{Lange2019}.  In this case,
the $P(k)$ shape is no longer fixed to that of {\it Planck} and
$\om{\neq}\omplk$. Accordingly, the value of $\bg^{\mathrm{clusters}}$ shifts
to $1.55{\pm}0.19$, but still disagrees strongly with the
clustering- and lensing-only estimates. Therefore, the existence of the
$\bg$-tension is robust against variations in the shape of
$P(k)$ or $\om$.

\begin{figure}
    \centering
    \includegraphics[width=1\linewidth]{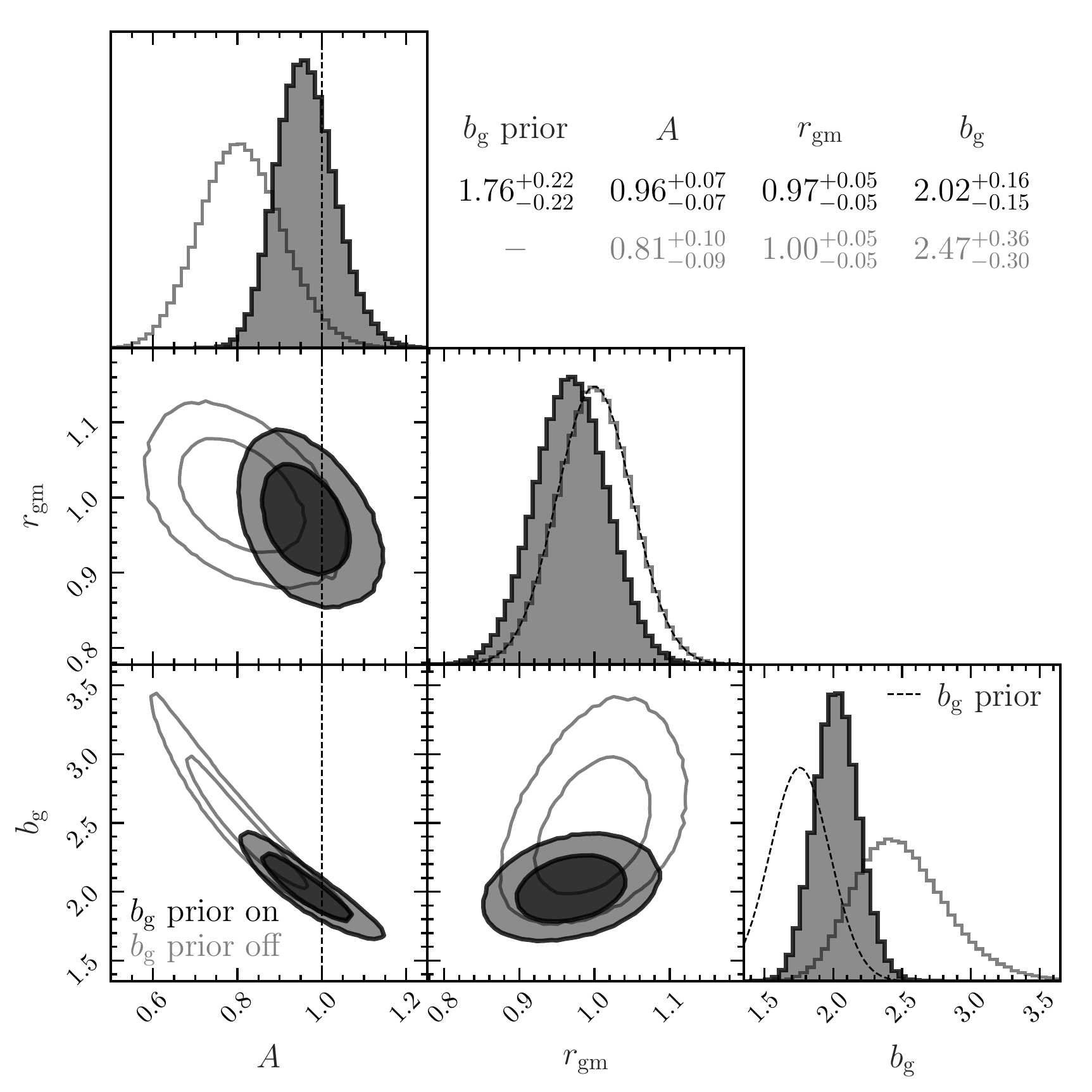}
    \caption{Constraints on the three parameters of the linear assessment
    model from our fiducial Bayesian analysis using the $\bg$
    prior~(dark filled contours/histograms). Light
    open contours/histograms show the constraints from the previous analysis
    without the $\bg$ prior.
    Contours in the off-diagonal panels indicate the 68\% and 95\%
	confidence regions on the 2D plane, while histograms in the
	diagonal panels show the 1D posterior distributions of individual
	parameters, with thin dashed curves in the $\bg$ and $\rgm$ panels
	representing the Gaussian priors. The two sets of $1\sigma$ posterior
	constraints are listed in the top right.}
    \label{fig:corner}
\end{figure}

Since the cluster-based bias measurement is independent of $A$ and derived
from external datasets, we can incorporate
$\bg^{\mathrm{clusters}}\eq1.76{\pm}0.22$ as a prior in the Bayesian
analysis using our linear assessment model developed in \S\ref{sec:simple}.
Figure~\ref{fig:corner} shows the two sets of posterior constraints on the
model parameters in both 1D~(diagonal) and 2D~(off-diagonal) parameter
spaces, one with the $\bg$-prior on~(dark filled histograms and contours) and
the other off~(light open). The contour levels are the 68\% and 95\% confidence
limits, with the thin dashed curve in the bottom right panel indicating our
$\bg$ prior, while the two sets of $1\sigma$ constraints are listed on the
top right.  The 1D posterior distribution of $\rgm$ coincides with the
Gaussian prior on $\rgm$~(thin dashed) when the $\bg$ prior is off, but
slides to $\rgm\eq0.97{\pm}0.05$ when the $\bg$ prior is on, consistent
with $\rcg\eq0.96{\pm}0.03$ and $\rcm{=}1$.  Meanwhile, our 1D posterior
constraint on $\bg$~($2.02_{-0.15}^{+0.16}$) is significantly lower than
the prior-off constraint~($2.47_{-0.30}^{+0.36}$), albeit slightly higher
than the prior.

Finally, in the bottom left panel of Figure~\ref{fig:corner}, we
successfully break the otherwise strong degeneracy between $A$ and $\bg$
with the help of the external prior, yielding a much better
agreement with the {\it Planck} cosmology~($A\eq1{\pm}{0.007}$; vertical
dashed line). After marginalizing over the uncertainties in $\bg$ and
$\rgm$, we derive a stringent 1D constraint of $A{=}0.96{\pm}{0.07}$,
finding no tension with {\it Planck}.

\section{Conclusion}
\label{sec:conc}

In this letter, we develop a simple method based on the linear
theory to assess the consistency between the clustering+lensing of BOSS
LOWZ galaxies and the {\it Planck} cosmology. Focusing on scales
$10{<}R{<}30\,\hmpc$ and assuming a fixed shape for the $\pklin$, we can
accurately model the clustering and lensing of galaxies using only three
parameters, $A$, $\bg$, and $\rgm$, where $A{=}1$ in the {\it Planck}
cosmology.

By examining the cross-correlation matrix of galaxy subsamples divided by
stellar mass, we discover that the LOWZ galaxies with $\log\ms{<}11.3$ are
affected by some unknown observational systematics that lead to spurious
correlations across large line-of-sight distances. We then build a clean
LOWZ galaxy sample~(C-LOWZ) by only including galaxies with
$\log\ms{>}11.3$. Applying our linear assessment model to the C-LOWZ
sample, we obtain a constraint of $A\eq0.81_{-0.09}^{+0.10}$, which
reproduces a $2\sigma$ tension with {\it Planck}.

However, there exists a strong degeneracy between our key parameter $A$ and
the nuisance parameter $\bg$. We develop a novel method of measuring $\bg$
by combining the cluster-galaxy cross-correlation and cluster weak lensing
using an overlapping sample of redMaPPer clusters. Intriguingly, despite
the discrepancy between the biases inferred from
clustering~($2.00{\pm}0.07$) and lensing~($1.63{\pm}0.17$) in {\it Planck},
both are consistent with our new cluster-based bias
estimate~($1.76{\pm}0.22$).  Applying this independent bias measurement as
a prior to our Bayesian analysis, we successfully break the degeneracy
between $A$ and $\bg$, and derive a stringent 1D posterior constraint of
$A\eq0.96_{-0.07}^{+0.07}$, in good agreement with the {\it Planck}
cosmology.

Our result suggests that the large-scale structure traced by the LOWZ
galaxies at $z{\sim}0.25$ is statistically consistent with {\it Planck},
while the clustering-lensing mismatch may be caused by some
observational systematics~(e.g., during the target selection). With the
upcoming flux-limited galaxy sample at the same redshift from the Dark
Energy Spectroscopic Instrument~\citep{DESI2022}, we will acquire a clearer
understanding of the clustering-lensing mismatch of the BOSS LOWZ sample,
whether it be a tension of $\Seight$ or $\bg$.



\begin{acknowledgments}
We thank Sukhdeep Singh, Zeyang Sun, and Jun Zhang for helpful discussions.
This work is supported by the National Key Basic Research and Development
Program of China (No. 2018YFA0404504, 2022YFF0503403), the National Science
Foundation of China (12173024, 11890692, 11621303, 11973070, 11873038), the
China Manned Space Project (No. CMS-CSST-2021-A01, CMS-CSST-2021-A02,
CMS-CSST-2021-B01), and the ``111'' project of the Ministry of Education
under grant No. B20019. Y.Z. acknowledges the generous sponsorship from
Yangyang Development Fund, and thanks Cathy Huang for her hospitality
during the pandemic.  H.S. acknowledges the Shanghai Committee of Science
and Technology grant No.19ZR1466600 and Key Research Program of Frontier
Sciences, CAS, Grant No. ZDBS-LY-7013.
\end{acknowledgments}

\bibliography{lowz}

\begin{thebibliography}{}
\expandafter\ifx\csname natexlab\endcsname\relax\def\natexlab#1{#1}\fi
\providecommand{\url}[1]{\href{#1}{#1}}
\providecommand{\dodoi}[1]{doi:~\href{http://doi.org/#1}{\nolinkurl{#1}}}
\providecommand{\doeprint}[1]{\href{http://ascl.net/#1}{\nolinkurl{http://ascl.net/#1}}}
\providecommand{\doarXiv}[1]{\href{https://arxiv.org/abs/#1}{\nolinkurl{https://arxiv.org/abs/#1}}}

\bibitem[{{Abareshi} {et~al.}(2022){Abareshi}, {Aguilar}, {Ahlen}, {Alam},
  {Alexander}, {Alfarsy}, {Allen}, {Allende Prieto}, {Alves}, {Ameel}, \&
  et~al.}]{DESI2022}
{Abareshi}, B., {Aguilar}, J., {Ahlen}, S., {et~al.} 2022, \aj, 164, 207,
  \dodoi{10.3847/1538-3881/ac882b}

\bibitem[{{Aihara} {et~al.}(2011){Aihara}, {Allende Prieto}, {An}, {Anderson},
  {Aubourg}, {Balbinot}, {Beers}, {Berlind}, {Bickerton}, {Bizyaev}, {Blanton},
  {Bochanski}, {Bolton}, {Bovy}, {Brand t}, {Brinkmann}, {Brown}, {Brownstein},
  {Busca}, {Campbell}, {Carr}, {Chen}, {Chiappini}, {Comparat}, {Connolly},
  {Cortes}, {Croft}, {Cuesta}, {da Costa}, {Davenport}, {Dawson}, {Dhital},
  {Ealet}, {Ebelke}, {Edmondson}, {Eisenstein}, {Escoffier}, {Esposito},
  {Evans}, {Fan}, {Femen{\'\i}a Castell{\'a}}, {Font-Ribera}, {Frinchaboy},
  {Ge}, {Gillespie}, {Gilmore}, {Gonz{\'a}lez Hern{\'a}ndez}, {Gott}, {Gould},
  {Grebel}, {Gunn}, {Hamilton}, {Harding}, {Harris}, {Hawley}, {Hearty}, {Ho},
  {Hogg}, {Holtzman}, {Honscheid}, {Inada}, {Ivans}, {Jiang}, {Johnson},
  {Jordan}, {Jordan}, {Kazin}, {Kirkby}, {Klaene}, {Knapp}, {Kneib},
  {Kochanek}, {Koesterke}, {Kollmeier}, {Kron}, {Lampeitl}, {Lang}, {Le Goff},
  {Lee}, {Lin}, {Long}, {Loomis}, {Lucatello}, {Lundgren}, {Lupton}, {Ma},
  {MacDonald}, {Mahadevan}, {Maia}, {Makler}, {Malanushenko}, {Malanushenko},
  {Mandelbaum}, {Maraston}, {Margala}, {Masters}, {McBride}, {McGehee},
  {McGreer}, {M{\'e}nard}, {Miralda-Escud{\'e}}, {Morrison}, {Mullally},
  {Muna}, {Munn}, {Murayama}, {Myers}, {Naugle}, {Neto}, {Nguyen}, {Nichol},
  {O'Connell}, {Ogando}, {Olmstead}, {Oravetz}, {Padmanabhan},
  {Palanque-Delabrouille}, {Pan}, {Pandey}, {P{\^a}ris}, {Percival},
  {Petitjean}, {Pfaffenberger}, {Pforr}, {Phleps}, {Pichon}, {Pieri}, {Prada},
  {Price-Whelan}, {Raddick}, {Ramos}, {Reyl{\'e}}, {Rich}, {Richards}, {Rix},
  {Robin}, {Rocha-Pinto}, {Rockosi}, {Roe}, {Rollinde}, {Ross}, {Ross},
  {Rossetto}, {S{\'a}nchez}, {Sayres}, {Schlegel}, {Schlesinger}, {Schmidt},
  {Schneider}, {Sheldon}, {Shu}, {Simmerer}, {Simmons}, {Sivarani}, {Snedden},
  {Sobeck}, {Steinmetz}, {Strauss}, {Szalay}, {Tanaka}, {Thakar}, {Thomas},
  {Tinker}, {Tofflemire}, {Tojeiro}, {Tremonti}, {Vandenberg}, {Vargas
  Maga{\~n}a}, {Verde}, {Vogt}, {Wake}, {Wang}, {Weaver}, {Weinberg}, {White},
  {White}, {Yanny}, {Yasuda}, {Yeche}, \& {Zehavi}}]{Aihara2011}
{Aihara}, H., {Allende Prieto}, C., {An}, D., {et~al.} 2011, \apjs, 193, 29,
  \dodoi{10.1088/0067-0049/193/2/29}

\bibitem[{{Alam} {et~al.}(2015){Alam}, {Albareti}, {Allende Prieto}, {Anders},
  {Anderson}, {Anderton}, {Andrews}, {Armengaud}, {Aubourg}, {Bailey}, \&
  et~al.}]{Alam2015}
{Alam}, S., {Albareti}, F.~D., {Allende Prieto}, C., {et~al.} 2015, \apjs, 219,
  12, \dodoi{10.1088/0067-0049/219/1/12}

\bibitem[{{Alam} {et~al.}(2021){Alam}, {Aubert}, {Avila}, {Balland},
  {Bautista}, {Bershady}, {Bizyaev}, {Blanton}, {Bolton}, {Bovy}, {Brinkmann},
  {Brownstein}, {Burtin}, {Chabanier}, {Chapman}, {Choi}, {Chuang}, {Comparat},
  {Cousinou}, {Cuceu}, {Dawson}, {de la Torre}, {de Mattia}, {Agathe}, {des
  Bourboux}, {Escoffier}, {Etourneau}, {Farr}, {Font-Ribera}, {Frinchaboy},
  {Fromenteau}, {Gil-Mar{\'\i}n}, {Le Goff}, {Gonzalez-Morales},
  {Gonzalez-Perez}, {Grabowski}, {Guy}, {Hawken}, {Hou}, {Kong}, {Parker},
  {Klaene}, {Kneib}, {Lin}, {Long}, {Lyke}, {de la Macorra}, {Martini},
  {Masters}, {Mohammad}, {Moon}, {Mueller}, {Mu{\~n}oz-Guti{\'e}rrez}, {Myers},
  {Nadathur}, {Neveux}, {Newman}, {Noterdaeme}, {Oravetz}, {Oravetz},
  {Palanque-Delabrouille}, {Pan}, {Paviot}, {Percival}, {P{\'e}rez-R{\`a}fols},
  {Petitjean}, {Pieri}, {Prakash}, {Raichoor}, {Ravoux}, {Rezaie}, {Rich},
  {Ross}, {Rossi}, {Ruggeri}, {Ruhlmann-Kleider}, {S{\'a}nchez}, {S{\'a}nchez},
  {S{\'a}nchez-Gallego}, {Sayres}, {Schneider}, {Seo}, {Shafieloo}, {Slosar},
  {Smith}, {Stermer}, {Tamone}, {Tinker}, {Tojeiro}, {Vargas-Maga{\~n}a},
  {Variu}, {Wang}, {Weaver}, {Weijmans}, {Y{\`e}che}, {Zarrouk}, {Zhao},
  {Zhao}, \& {Zheng}}]{Alam2021}
{Alam}, S., {Aubert}, M., {Avila}, S., {et~al.} 2021, \prd, 103, 083533,
  \dodoi{10.1103/PhysRevD.103.083533}

\bibitem[{{Amon} \& {Efstathiou}(2022)}]{Amon2022lows8nonlinear}
{Amon}, A., \& {Efstathiou}, G. 2022, \mnras, 516, 5355,
  \dodoi{10.1093/mnras/stac2429}

\bibitem[{{Amon} {et~al.}(2022){Amon}, {Gruen}, {Troxel}, {MacCrann},
  {Dodelson}, {Choi}, {Doux}, {Secco}, {Samuroff}, {Krause}, {Cordero},
  {Myles}, {DeRose}, {Wechsler}, {Gatti}, {Navarro-Alsina}, {Bernstein},
  {Jain}, {Blazek}, {Alarcon}, {Fert{\'e}}, {Lemos}, {Raveri}, {Campos},
  {Prat}, {S{\'a}nchez}, {Jarvis}, {Alves}, {Andrade-Oliveira}, {Baxter},
  {Bechtol}, {Becker}, {Bridle}, {Camacho}, {Carnero Rosell}, {Carrasco Kind},
  {Cawthon}, {Chang}, {Chen}, {Chintalapati}, {Crocce}, {Davis}, {Diehl},
  {Drlica-Wagner}, {Eckert}, {Eifler}, {Elvin-Poole}, {Everett}, {Fang},
  {Fosalba}, {Friedrich}, {Gaztanaga}, {Giannini}, {Gruendl}, {Harrison},
  {Hartley}, {Herner}, {Huang}, {Huff}, {Huterer}, {Kuropatkin}, {Leget},
  {Liddle}, {McCullough}, {Muir}, {Pandey}, {Park}, {Porredon}, {Refregier},
  {Rollins}, {Roodman}, {Rosenfeld}, {Ross}, {Rykoff}, {Sanchez},
  {Sevilla-Noarbe}, {Sheldon}, {Shin}, {Troja}, {Tutusaus}, {Tutusaus},
  {Varga}, {Weaverdyck}, {Yanny}, {Yin}, {Zhang}, {Zuntz}, {Aguena}, {Allam},
  {Annis}, {Bacon}, {Bertin}, {Bhargava}, {Brooks}, {Buckley-Geer}, {Burke},
  {Carretero}, {Costanzi}, {da Costa}, {Pereira}, {De Vicente}, {Desai},
  {Dietrich}, {Doel}, {Ferrero}, {Flaugher}, {Frieman}, {Garc{\'\i}a-Bellido},
  {Gaztanaga}, {Gerdes}, {Giannantonio}, {Gschwend}, {Gutierrez}, {Hinton},
  {Hollowood}, {Honscheid}, {Hoyle}, {James}, {Kron}, {Kuehn}, {Lahav}, {Lima},
  {Lin}, {Maia}, {Marshall}, {Martini}, {Melchior}, {Menanteau}, {Miquel},
  {Mohr}, {Morgan}, {Ogando}, {Palmese}, {Paz-Chinch{\'o}n}, {Petravick},
  {Pieres}, {Romer}, {Sanchez}, {Scarpine}, {Schubnell}, {Serrano}, {Smith},
  {Soares-Santos}, {Tarle}, {Thomas}, {To}, {Weller}, \& {DES
  Collaboration}}]{Amon2022DES}
{Amon}, A., {Gruen}, D., {Troxel}, M.~A., {et~al.} 2022, \prd, 105, 023514,
  \dodoi{10.1103/PhysRevD.105.023514}

\bibitem[{{Amon} {et~al.}(2023){Amon}, {Robertson}, {Miyatake}, {Heymans},
  {White}, {DeRose}, {Yuan}, {Wechsler}, {Varga}, {Bocquet}, {Dvornik}, {More},
  {Ross}, {Hoekstra}, {Alarcon}, {Asgari}, {Blazek}, {Campos}, {Chen}, {Choi},
  {Crocce}, {Diehl}, {Doux}, {Eckert}, {Elvin-Poole}, {Everett}, {Fert{\'e}},
  {Gatti}, {Giannini}, {Gruen}, {Gruendl}, {Hartley}, {Herner}, {Hildebrandt},
  {Huang}, {Huff}, {Joachimi}, {Lee}, {MacCrann}, {Myles}, {Navarro-Alsina},
  {Nishimichi}, {Prat}, {Secco}, {Sevilla-Noarbe}, {Sheldon}, {Shin},
  {Tr{\"o}ster}, {Troxel}, {Tutusaus}, {Wright}, {Yin}, {Aguena}, {Allam},
  {Annis}, {Bacon}, {Bilicki}, {Brooks}, {Burke}, {Carnero Rosell},
  {Carretero}, {Castander}, {Cawthon}, {Costanzi}, {da Costa}, {Pereira}, {de
  Jong}, {De Vicente}, {Desai}, {Dietrich}, {Doel}, {Ferrero}, {Frieman},
  {Garc{\'\i}a-Bellido}, {Gerdes}, {Gschwend}, {Gutierrez}, {Hinton},
  {Hollowood}, {Honscheid}, {Huterer}, {Kannawadi}, {Kuehn}, {Kuropatkin},
  {Lahav}, {Lima}, {Maia}, {Marshall}, {Menanteau}, {Miquel}, {Mohr}, {Morgan},
  {Muir}, {Paz-Chinch{\'o}n}, {Pieres}, {Plazas Malag{\'o}n}, {Porredon},
  {Rodriguez-Monroy}, {Roodman}, {Sanchez}, {Serrano}, {Shan}, {Suchyta},
  {Swanson}, {Tarle}, {Thomas}, {To}, \& {Zhang}}]{Amon2023}
{Amon}, A., {Robertson}, N.~C., {Miyatake}, H., {et~al.} 2023, \mnras, 518,
  477, \dodoi{10.1093/mnras/stac2938}

\bibitem[{{Asgari} {et~al.}(2021){Asgari}, {Lin}, {Joachimi}, {Giblin},
  {Heymans}, {Hildebrandt}, {Kannawadi}, {St{\"o}lzner}, {Tr{\"o}ster}, {van
  den Busch}, {Wright}, {Bilicki}, {Blake}, {de Jong}, {Dvornik}, {Erben},
  {Getman}, {Hoekstra}, {K{\"o}hlinger}, {Kuijken}, {Miller}, {Radovich},
  {Schneider}, {Shan}, \& {Valentijn}}]{Asgari2021KiDS}
{Asgari}, M., {Lin}, C.-A., {Joachimi}, B., {et~al.} 2021, \aap, 645, A104,
  \dodoi{10.1051/0004-6361/202039070}

\bibitem[{{Baldauf} {et~al.}(2010){Baldauf}, {Smith}, {Seljak}, \&
  {Mandelbaum}}]{Baldauf2010Upsilon}
{Baldauf}, T., {Smith}, R.~E., {Seljak}, U., \& {Mandelbaum}, R. 2010, \prd,
  81, 063531, \dodoi{10.1103/PhysRevD.81.063531}

\bibitem[{{Beltz-Mohrmann} {et~al.}(2022){Beltz-Mohrmann}, {Szewciw},
  {Berlind}, \& {Sinha}}]{Beltz-Mohrmann2022}
{Beltz-Mohrmann}, G.~D., {Szewciw}, A.~O., {Berlind}, A.~A., \& {Sinha}, M.
  2022, arXiv e-prints, arXiv:2211.16105, \dodoi{10.48550/arXiv.2211.16105}

\bibitem[{{Cacciato} {et~al.}(2012){Cacciato}, {Lahav}, {van den Bosch},
  {Hoekstra}, \& {Dekel}}]{Cacciato2012}
{Cacciato}, M., {Lahav}, O., {van den Bosch}, F.~C., {Hoekstra}, H., \&
  {Dekel}, A. 2012, \mnras, 426, 566, \dodoi{10.1111/j.1365-2966.2012.21762.x}

\bibitem[{{Chaves-Montero} {et~al.}(2022){Chaves-Montero}, {Angulo}, \&
  {Contreras}}]{ChavesMontero2022}
{Chaves-Montero}, J., {Angulo}, R.~E., \& {Contreras}, S. 2022, arXiv e-prints,
  arXiv:2211.01744.
\newblock \doarXiv{2211.01744}

\bibitem[{{Chen} {et~al.}(2012){Chen}, {Kauffmann}, {Tremonti}, {White},
  {Heckman}, {Kova{\v{c}}}, {Bundy}, {Chisholm}, {Maraston}, {Schneider},
  {Bolton}, {Weaver}, \& {Brinkmann}}]{Chen2012StellarMass}
{Chen}, Y.-M., {Kauffmann}, G., {Tremonti}, C.~A., {et~al.} 2012, \mnras, 421,
  314, \dodoi{10.1111/j.1365-2966.2011.20306.x}

\bibitem[{{Contreras} {et~al.}(2022){Contreras}, {Angulo}, {Chaves-Montero},
  {White}, \& {Aric{\`o}}}]{Contreras2022consistentGCGGL}
{Contreras}, S., {Angulo}, R.~E., {Chaves-Montero}, J., {White}, S. D.~M., \&
  {Aric{\`o}}, G. 2022, arXiv e-prints, arXiv:2211.11745.
\newblock \doarXiv{2211.11745}

\bibitem[{{Dawson} {et~al.}(2013){Dawson}, {Schlegel}, {Ahn}, {Anderson},
  {Aubourg}, {Bailey}, {Barkhouser}, {Bautista}, {Beifiori}, {Berlind},
  {Bhardwaj}, {Bizyaev}, {Blake}, {Blanton}, {Blomqvist}, {Bolton}, {Borde},
  {Bovy}, {Brandt}, {Brewington}, {Brinkmann}, {Brown}, {Brownstein}, {Bundy},
  {Busca}, {Carithers}, {Carnero}, {Carr}, {Chen}, {Comparat}, {Connolly},
  {Cope}, {Croft}, {Cuesta}, {da Costa}, {Davenport}, {Delubac}, {de Putter},
  {Dhital}, {Ealet}, {Ebelke}, {Eisenstein}, {Escoffier}, {Fan}, {Filiz Ak},
  {Finley}, {Font-Ribera}, {G{\'e}nova-Santos}, {Gunn}, {Guo}, {Haggard},
  {Hall}, {Hamilton}, {Harris}, {Harris}, {Ho}, {Hogg}, {Holder}, {Honscheid},
  {Huehnerhoff}, {Jordan}, {Jordan}, {Kauffmann}, {Kazin}, {Kirkby}, {Klaene},
  {Kneib}, {Le Goff}, {Lee}, {Long}, {Loomis}, {Lundgren}, {Lupton}, {Maia},
  {Makler}, {Malanushenko}, {Malanushenko}, {Mandelbaum}, {Manera}, {Maraston},
  {Margala}, {Masters}, {McBride}, {McDonald}, {McGreer}, {McMahon}, {Mena},
  {Miralda-Escud{\'e}}, {Montero-Dorta}, {Montesano}, {Muna}, {Myers},
  {Naugle}, {Nichol}, {Noterdaeme}, {Nuza}, {Olmstead}, {Oravetz}, {Oravetz},
  {Owen}, {Padmanabhan}, {Palanque-Delabrouille}, {Pan}, {Parejko},
  {P{\^a}ris}, {Percival}, {P{\'e}rez-Fournon}, {P{\'e}rez-R{\`a}fols},
  {Petitjean}, {Pfaffenberger}, {Pforr}, {Pieri}, {Prada}, {Price-Whelan},
  {Raddick}, {Rebolo}, {Rich}, {Richards}, {Rockosi}, {Roe}, {Ross}, {Ross},
  {Rossi}, {Rubi{\~n}o-Martin}, {Samushia}, {S{\'a}nchez}, {Sayres}, {Schmidt},
  {Schneider}, {Sc{\'o}ccola}, {Seo}, {Shelden}, {Sheldon}, {Shen}, {Shu},
  {Slosar}, {Smee}, {Snedden}, {Stauffer}, {Steele}, {Strauss}, {Streblyanska},
  {Suzuki}, {Swanson}, {Tal}, {Tanaka}, {Thomas}, {Tinker}, {Tojeiro},
  {Tremonti}, {Vargas Maga{\~n}a}, {Verde}, {Viel}, {Wake}, {Watson}, {Weaver},
  {Weinberg}, {Weiner}, {West}, {White}, {Wood-Vasey}, {Yeche}, {Zehavi},
  {Zhao}, \& {Zheng}}]{Dawson2013}
{Dawson}, K.~S., {Schlegel}, D.~J., {Ahn}, C.~P., {et~al.} 2013, \aj, 145, 10,
  \dodoi{10.1088/0004-6256/145/1/10}

\bibitem[{{Dey} {et~al.}(2019){Dey}, {Schlegel}, {Lang}, {Blum}, {Burleigh},
  {Fan}, {Findlay}, {Finkbeiner}, {Herrera}, {Juneau}, {Landriau}, {Levi},
  {McGreer}, {Meisner}, {Myers}, {Moustakas}, {Nugent}, {Patej}, {Schlafly},
  {Walker}, {Valdes}, {Weaver}, {Y{\`e}che}, {Zou}, {Zhou}, {Abareshi},
  {Abbott}, {Abolfathi}, {Aguilera}, {Alam}, {Allen}, {Alvarez}, {Annis},
  {Ansarinejad}, {Aubert}, {Beechert}, {Bell}, {BenZvi}, {Beutler}, {Bielby},
  {Bolton}, {Brice{\~n}o}, {Buckley-Geer}, {Butler}, {Calamida}, {Carlberg},
  {Carter}, {Casas}, {Castander}, {Choi}, {Comparat}, {Cukanovaite}, {Delubac},
  {DeVries}, {Dey}, {Dhungana}, {Dickinson}, {Ding}, {Donaldson}, {Duan},
  {Duckworth}, {Eftekharzadeh}, {Eisenstein}, {Etourneau}, {Fagrelius},
  {Farihi}, {Fitzpatrick}, {Font-Ribera}, {Fulmer}, {G{\"a}nsicke},
  {Gaztanaga}, {George}, {Gerdes}, {Gontcho}, {Gorgoni}, {Green}, {Guy},
  {Harmer}, {Hernandez}, {Honscheid}, {Huang}, {James}, {Jannuzi}, {Jiang},
  {Joyce}, {Karcher}, {Karkar}, {Kehoe}, {Kneib}, {Kueter-Young}, {Lan},
  {Lauer}, {Le Guillou}, {Le Van Suu}, {Lee}, {Lesser}, {Perreault Levasseur},
  {Li}, {Mann}, {Marshall}, {Mart{\'\i}nez-V{\'a}zquez}, {Martini}, {du Mas des
  Bourboux}, {McManus}, {Meier}, {M{\'e}nard}, {Metcalfe},
  {Mu{\~n}oz-Guti{\'e}rrez}, {Najita}, {Napier}, {Narayan}, {Newman}, {Nie},
  {Nord}, {Norman}, {Olsen}, {Paat}, {Palanque-Delabrouille}, {Peng},
  {Poppett}, {Poremba}, {Prakash}, {Rabinowitz}, {Raichoor}, {Rezaie},
  {Robertson}, {Roe}, {Ross}, {Ross}, {Rudnick}, {Safonova}, {Saha},
  {S{\'a}nchez}, {Savary}, {Schweiker}, {Scott}, {Seo}, {Shan}, {Silva},
  {Slepian}, {Soto}, {Sprayberry}, {Staten}, {Stillman}, {Stupak}, {Summers},
  {Sien Tie}, {Tirado}, {Vargas-Maga{\~n}a}, {Vivas}, {Wechsler}, {Williams},
  {Yang}, {Yang}, {Yapici}, {Zaritsky}, {Zenteno}, {Zhang}, {Zhang}, {Zhou}, \&
  {Zhou}}]{Dey2019}
{Dey}, A., {Schlegel}, D.~J., {Lang}, D., {et~al.} 2019, \aj, 157, 168,
  \dodoi{10.3847/1538-3881/ab089d}

\bibitem[{{Eisenstein} {et~al.}(2011){Eisenstein}, {Weinberg}, {Agol},
  {Aihara}, {Allende Prieto}, {Anderson}, {Arns}, {Aubourg}, {Bailey},
  {Balbinot}, \& et~al.}]{Eisenstein2011}
{Eisenstein}, D.~J., {Weinberg}, D.~H., {Agol}, E., {et~al.} 2011, \aj, 142,
  72, \dodoi{10.1088/0004-6256/142/3/72}

\bibitem[{{Guo} {et~al.}(2018){Guo}, {Yang}, \& {Lu}}]{Guo2018CSMF}
{Guo}, H., {Yang}, X., \& {Lu}, Y. 2018, \apj, 858, 30,
  \dodoi{10.3847/1538-4357/aabc56}

\bibitem[{{Huterer}(2022)}]{Huterer2022}
{Huterer}, D. 2022, arXiv e-prints, arXiv:2212.05003,
  \dodoi{10.48550/arXiv.2212.05003}

\bibitem[{{Landy} \& {Szalay}(1993)}]{Landy1993LSEstimator}
{Landy}, S.~D., \& {Szalay}, A.~S. 1993, \apj, 412, 64, \dodoi{10.1086/172900}

\bibitem[{{Lange} {et~al.}(2021){Lange}, {Leauthaud}, {Singh}, {Guo}, {Zhou},
  {Smith}, \& {Cyr-Racine}}]{Lange2021}
{Lange}, J.~U., {Leauthaud}, A., {Singh}, S., {et~al.} 2021, \mnras, 502, 2074,
  \dodoi{10.1093/mnras/stab189}

\bibitem[{{Lange} {et~al.}(2019){Lange}, {Yang}, {Guo}, {Luo}, \& {van den
  Bosch}}]{Lange2019}
{Lange}, J.~U., {Yang}, X., {Guo}, H., {Luo}, W., \& {van den Bosch}, F.~C.
  2019, \mnras, 488, 5771, \dodoi{10.1093/mnras/stz2124}

\bibitem[{{Leauthaud} {et~al.}(2017){Leauthaud}, {Saito}, {Hilbert},
  {Barreira}, {More}, {White}, {Alam}, {Behroozi}, {Bundy}, {Coupon}, {Erben},
  {Heymans}, {Hildebrandt}, {Mandelbaum}, {Miller}, {Moraes}, {Pereira},
  {Rodr{\'\i}guez-Torres}, {Schmidt}, {Shan}, {Viel}, \&
  {Villaescusa-Navarro}}]{Leauthaud2017}
{Leauthaud}, A., {Saito}, S., {Hilbert}, S., {et~al.} 2017, \mnras, 467, 3024,
  \dodoi{10.1093/mnras/stx258}

\bibitem[{{Mandelbaum} {et~al.}(2013){Mandelbaum}, {Slosar}, {Baldauf},
  {Seljak}, {Hirata}, {Nakajima}, {Reyes}, \& {Smith}}]{Mandelbaum2013}
{Mandelbaum}, R., {Slosar}, A., {Baldauf}, T., {et~al.} 2013, \mnras, 432,
  1544, \dodoi{10.1093/mnras/stt572}

\bibitem[{{More} {et~al.}(2015){More}, {Miyatake}, {Mandelbaum}, {Takada},
  {Spergel}, {Brownstein}, \& {Schneider}}]{More2015}
{More}, S., {Miyatake}, H., {Mandelbaum}, R., {et~al.} 2015, \apj, 806, 2,
  \dodoi{10.1088/0004-637X/806/1/2}

\bibitem[{{Planck Collaboration} {et~al.}(2020){Planck Collaboration},
  {Aghanim}, {Akrami}, {Ashdown}, {Aumont}, {Baccigalupi}, {Ballardini},
  {Banday}, {Barreiro}, {Bartolo}, {Basak}, {Battye}, {Benabed}, {Bernard},
  {Bersanelli}, {Bielewicz}, {Bock}, {Bond}, {Borrill}, {Bouchet}, {Boulanger},
  {Bucher}, {Burigana}, {Butler}, {Calabrese}, {Cardoso}, {Carron},
  {Challinor}, {Chiang}, {Chluba}, {Colombo}, {Combet}, {Contreras}, {Crill},
  {Cuttaia}, {de Bernardis}, {de Zotti}, {Delabrouille}, {Delouis}, {Di
  Valentino}, {Diego}, {Dor{\'e}}, {Douspis}, {Ducout}, {Dupac}, {Dusini},
  {Efstathiou}, {Elsner}, {En{\ss}lin}, {Eriksen}, {Fantaye}, {Farhang},
  {Fergusson}, {Fernandez-Cobos}, {Finelli}, {Forastieri}, {Frailis},
  {Fraisse}, {Franceschi}, {Frolov}, {Galeotta}, {Galli}, {Ganga},
  {G{\'e}nova-Santos}, {Gerbino}, {Ghosh}, {Gonz{\'a}lez-Nuevo}, {G{\'o}rski},
  {Gratton}, {Gruppuso}, {Gudmundsson}, {Hamann}, {Handley}, {Hansen},
  {Herranz}, {Hildebrandt}, {Hivon}, {Huang}, {Jaffe}, {Jones}, {Karakci},
  {Keih{\"a}nen}, {Keskitalo}, {Kiiveri}, {Kim}, {Kisner}, {Knox},
  {Krachmalnicoff}, {Kunz}, {Kurki-Suonio}, {Lagache}, {Lamarre}, {Lasenby},
  {Lattanzi}, {Lawrence}, {Le Jeune}, {Lemos}, {Lesgourgues}, {Levrier},
  {Lewis}, {Liguori}, {Lilje}, {Lilley}, {Lindholm}, {L{\'o}pez-Caniego},
  {Lubin}, {Ma}, {Mac{\'\i}as-P{\'e}rez}, {Maggio}, {Maino}, {Mandolesi},
  {Mangilli}, {Marcos-Caballero}, {Maris}, {Martin}, {Martinelli},
  {Mart{\'\i}nez-Gonz{\'a}lez}, {Matarrese}, {Mauri}, {McEwen}, {Meinhold},
  {Melchiorri}, {Mennella}, {Migliaccio}, {Millea}, {Mitra},
  {Miville-Desch{\^e}nes}, {Molinari}, {Montier}, {Morgante}, {Moss}, {Natoli},
  {N{\o}rgaard-Nielsen}, {Pagano}, {Paoletti}, {Partridge}, {Patanchon},
  {Peiris}, {Perrotta}, {Pettorino}, {Piacentini}, {Polastri}, {Polenta},
  {Puget}, {Rachen}, {Reinecke}, {Remazeilles}, {Renzi}, {Rocha}, {Rosset},
  {Roudier}, {Rubi{\~n}o-Mart{\'\i}n}, {Ruiz-Granados}, {Salvati}, {Sandri},
  {Savelainen}, {Scott}, {Shellard}, {Sirignano}, {Sirri}, {Spencer},
  {Sunyaev}, {Suur-Uski}, {Tauber}, {Tavagnacco}, {Tenti}, {Toffolatti},
  {Tomasi}, {Trombetti}, {Valenziano}, {Valiviita}, {Van Tent}, {Vibert},
  {Vielva}, {Villa}, {Vittorio}, {Wandelt}, {Wehus}, {White}, {White},
  {Zacchei}, \& {Zonca}}]{Planck2020parameters}
{Planck Collaboration}, {Aghanim}, N., {Akrami}, Y., {et~al.} 2020, \aap, 641,
  A6, \dodoi{10.1051/0004-6361/201833910}

\bibitem[{{Reid} {et~al.}(2016){Reid}, {Ho}, {Padmanabhan}, {Percival},
  {Tinker}, {Tojeiro}, {White}, {Eisenstein}, {Maraston}, {Ross},
  {S{\'a}nchez}, {Schlegel}, {Sheldon}, {Strauss}, {Thomas}, {Wake}, {Beutler},
  {Bizyaev}, {Bolton}, {Brownstein}, {Chuang}, {Dawson}, {Harding}, {Kitaura},
  {Leauthaud}, {Masters}, {McBride}, {More}, {Olmstead}, {Oravetz}, {Nuza},
  {Pan}, {Parejko}, {Pforr}, {Prada}, {Rodr{\'\i}guez-Torres},
  {Salazar-Albornoz}, {Samushia}, {Schneider}, {Sc{\'o}ccola}, {Simmons}, \&
  {Vargas-Magana}}]{Reid2016}
{Reid}, B., {Ho}, S., {Padmanabhan}, N., {et~al.} 2016, \mnras, 455, 1553,
  \dodoi{10.1093/mnras/stv2382}

\bibitem[{{Reyes} {et~al.}(2012){Reyes}, {Mandelbaum}, {Gunn}, {Nakajima},
  {Seljak}, \& {Hirata}}]{Reyes2012}
{Reyes}, R., {Mandelbaum}, R., {Gunn}, J.~E., {et~al.} 2012, \mnras, 425, 2610,
  \dodoi{10.1111/j.1365-2966.2012.21472.x}

\bibitem[{{Rykoff} {et~al.}(2014){Rykoff}, {Rozo}, {Busha}, {Cunha},
  {Finoguenov}, {Evrard}, {Hao}, {Koester}, {Leauthaud}, {Nord}, {Pierre},
  {Reddick}, {Sadibekova}, {Sheldon}, \& {Wechsler}}]{Rykoff2014redmapper}
{Rykoff}, E.~S., {Rozo}, E., {Busha}, M.~T., {et~al.} 2014, \apj, 785, 104,
  \dodoi{10.1088/0004-637X/785/2/104}

\bibitem[{{Rykoff} {et~al.}(2016){Rykoff}, {Rozo}, {Hollowood},
  {Bermeo-Hernandez}, {Jeltema}, {Mayers}, {Romer}, {Rooney}, {Saro}, {Vergara
  Cervantes}, {Wechsler}, {Wilcox}, {Abbott}, {Abdalla}, {Allam}, {Annis},
  {Benoit-L{\'e}vy}, {Bernstein}, {Bertin}, {Brooks}, {Burke}, {Capozzi},
  {Carnero Rosell}, {Carrasco Kind}, {Castander}, {Childress}, {Collins},
  {Cunha}, {D'Andrea}, {da Costa}, {Davis}, {Desai}, {Diehl}, {Dietrich},
  {Doel}, {Evrard}, {Finley}, {Flaugher}, {Fosalba}, {Frieman}, {Glazebrook},
  {Goldstein}, {Gruen}, {Gruendl}, {Gutierrez}, {Hilton}, {Honscheid}, {Hoyle},
  {James}, {Kay}, {Kuehn}, {Kuropatkin}, {Lahav}, {Lewis}, {Lidman}, {Lima},
  {Maia}, {Mann}, {Marshall}, {Martini}, {Melchior}, {Miller}, {Miquel},
  {Mohr}, {Nichol}, {Nord}, {Ogando}, {Plazas}, {Reil}, {Sahl{\'e}n},
  {Sanchez}, {Santiago}, {Scarpine}, {Schubnell}, {Sevilla-Noarbe}, {Smith},
  {Soares-Santos}, {Sobreira}, {Stott}, {Suchyta}, {Swanson}, {Tarle},
  {Thomas}, {Tucker}, {Uddin}, {Viana}, {Vikram}, {Walker}, {Zhang}, \& {DES
  Collaboration}}]{Rykoff2016redmapper}
{Rykoff}, E.~S., {Rozo}, E., {Hollowood}, D., {et~al.} 2016, \apjs, 224, 1,
  \dodoi{10.3847/0067-0049/224/1/1}

\bibitem[{{Salcedo} {et~al.}(2022){Salcedo}, {Zu}, {Zhang}, {Wang}, {Yang},
  {Wu}, {Jing}, {Mo}, \& {Weinberg}}]{Salcedo2022}
{Salcedo}, A.~N., {Zu}, Y., {Zhang}, Y., {et~al.} 2022, Science China Physics,
  Mechanics, and Astronomy, 65, 109811, \dodoi{10.1007/s11433-022-1955-7}

\bibitem[{{Scolnic} {et~al.}(2018){Scolnic}, {Jones}, {Rest}, {Pan},
  {Chornock}, {Foley}, {Huber}, {Kessler}, {Narayan}, {Riess}, {Rodney},
  {Berger}, {Brout}, {Challis}, {Drout}, {Finkbeiner}, {Lunnan}, {Kirshner},
  {Sanders}, {Schlafly}, {Smartt}, {Stubbs}, {Tonry}, {Wood-Vasey}, {Foley},
  {Hand}, {Johnson}, {Burgett}, {Chambers}, {Draper}, {Hodapp}, {Kaiser},
  {Kudritzki}, {Magnier}, {Metcalfe}, {Bresolin}, {Gall}, {Kotak}, {McCrum}, \&
  {Smith}}]{Scolnic2018}
{Scolnic}, D.~M., {Jones}, D.~O., {Rest}, A., {et~al.} 2018, \apj, 859, 101,
  \dodoi{10.3847/1538-4357/aab9bb}

\bibitem[{{Secco} {et~al.}(2022){Secco}, {Samuroff}, {Krause}, {Jain},
  {Blazek}, {Raveri}, {Campos}, {Amon}, {Chen}, {Doux}, {Choi}, {Gruen},
  {Bernstein}, {Chang}, {DeRose}, {Myles}, {Fert{\'e}}, {Lemos}, {Huterer},
  {Prat}, {Troxel}, {MacCrann}, {Liddle}, {Kacprzak}, {Fang}, {S{\'a}nchez},
  {Pandey}, {Dodelson}, {Chintalapati}, {Hoffmann}, {Alarcon}, {Alves},
  {Andrade-Oliveira}, {Baxter}, {Bechtol}, {Becker}, {Brandao-Souza},
  {Camacho}, {Carnero Rosell}, {Carrasco Kind}, {Cawthon}, {Cordero}, {Crocce},
  {Davis}, {Di Valentino}, {Drlica-Wagner}, {Eckert}, {Eifler}, {Elidaiana},
  {Elsner}, {Elvin-Poole}, {Everett}, {Fosalba}, {Friedrich}, {Gatti},
  {Giannini}, {Gruendl}, {Harrison}, {Hartley}, {Herner}, {Huang}, {Huff},
  {Jarvis}, {Jeffrey}, {Kuropatkin}, {Leget}, {Muir}, {Mccullough}, {Navarro
  Alsina}, {Omori}, {Park}, {Porredon}, {Rollins}, {Roodman}, {Rosenfeld},
  {Ross}, {Rykoff}, {Sanchez}, {Sevilla-Noarbe}, {Sheldon}, {Shin}, {Troja},
  {Tutusaus}, {Varga}, {Weaverdyck}, {Wechsler}, {Yanny}, {Yin}, {Zhang},
  {Zuntz}, {Abbott}, {Aguena}, {Allam}, {Annis}, {Bacon}, {Bertin}, {Bhargava},
  {Bridle}, {Brooks}, {Buckley-Geer}, {Burke}, {Carretero}, {Costanzi}, {da
  Costa}, {De Vicente}, {Diehl}, {Dietrich}, {Doel}, {Ferrero}, {Flaugher},
  {Frieman}, {Garc{\'\i}a-Bellido}, {Gaztanaga}, {Gerdes}, {Giannantonio},
  {Gschwend}, {Gutierrez}, {Hinton}, {Hollowood}, {Honscheid}, {Hoyle},
  {James}, {Jeltema}, {Kuehn}, {Lahav}, {Lima}, {Lin}, {Maia}, {Marshall},
  {Martini}, {Melchior}, {Menanteau}, {Miquel}, {Mohr}, {Morgan}, {Ogando},
  {Palmese}, {Paz-Chinch{\'o}n}, {Petravick}, {Pieres}, {Plazas Malag{\'o}n},
  {Rodriguez-Monroy}, {Romer}, {Sanchez}, {Scarpine}, {Schubnell}, {Scolnic},
  {Serrano}, {Smith}, {Soares-Santos}, {Suchyta}, {Swanson}, {Tarle}, {Thomas},
  {To}, \& {DES Collaboration}}]{Secco2022DES}
{Secco}, L.~F., {Samuroff}, S., {Krause}, E., {et~al.} 2022, \prd, 105, 023515,
  \dodoi{10.1103/PhysRevD.105.023515}

\bibitem[{{Singh} {et~al.}(2020){Singh}, {Mandelbaum}, {Seljak},
  {Rodr{\'\i}guez-Torres}, \& {Slosar}}]{Singh2020}
{Singh}, S., {Mandelbaum}, R., {Seljak}, U., {Rodr{\'\i}guez-Torres}, S., \&
  {Slosar}, A. 2020, \mnras, 491, 51, \dodoi{10.1093/mnras/stz2922}

\bibitem[{{Takahashi} {et~al.}(2012){Takahashi}, {Sato}, {Nishimichi},
  {Taruya}, \& {Oguri}}]{Takahashi2012}
{Takahashi}, R., {Sato}, M., {Nishimichi}, T., {Taruya}, A., \& {Oguri}, M.
  2012, \apj, 761, 152, \dodoi{10.1088/0004-637X/761/2/152}

\bibitem[{{Tegmark} \& {Bromley}(1999)}]{Tegmark1999}
{Tegmark}, M., \& {Bromley}, B.~C. 1999, \apjl, 518, L69,
  \dodoi{10.1086/312068}

\bibitem[{{van den Bosch} {et~al.}(2013){van den Bosch}, {More}, {Cacciato},
  {Mo}, \& {Yang}}]{vandenBosch2013rrsd}
{van den Bosch}, F.~C., {More}, S., {Cacciato}, M., {Mo}, H., \& {Yang}, X.
  2013, \mnras, 430, 725, \dodoi{10.1093/mnras/sts006}

\bibitem[{{Weinberg} {et~al.}(2013){Weinberg}, {Mortonson}, {Eisenstein},
  {Hirata}, {Riess}, \& {Rozo}}]{Weinberg2013}
{Weinberg}, D.~H., {Mortonson}, M.~J., {Eisenstein}, D.~J., {et~al.} 2013,
  \physrep, 530, 87, \dodoi{10.1016/j.physrep.2013.05.001}

\bibitem[{{Wibking} {et~al.}(2020){Wibking}, {Weinberg}, {Salcedo}, {Wu},
  {Singh}, {Rodr{\'\i}guez-Torres}, {Garrison}, \& {Eisenstein}}]{Wibking2020}
{Wibking}, B.~D., {Weinberg}, D.~H., {Salcedo}, A.~N., {et~al.} 2020, \mnras,
  492, 2872, \dodoi{10.1093/mnras/stz3423}

\bibitem[{{Yuan} {et~al.}(2020){Yuan}, {Eisenstein}, \&
  {Leauthaud}}]{Yuan2020MNRASAssmeblyBias}
{Yuan}, S., {Eisenstein}, D.~J., \& {Leauthaud}, A. 2020, \mnras, 493, 5551,
  \dodoi{10.1093/mnras/staa634}

\bibitem[{{Zu}(2020)}]{Zu2020}
{Zu}, Y. 2020, arXiv e-prints, arXiv:2010.01143.
\newblock \doarXiv{2010.01143}

\bibitem[{{Zu} \& {Mandelbaum}(2015)}]{Zu2015}
{Zu}, Y., \& {Mandelbaum}, R. 2015, \mnras, 454, 1161,
  \dodoi{10.1093/mnras/stv2062}

\bibitem[{{Zu} {et~al.}(2014){Zu}, {Weinberg}, {Rozo}, {Sheldon}, {Tinker}, \&
  {Becker}}]{Zu2014}
{Zu}, Y., {Weinberg}, D.~H., {Rozo}, E., {et~al.} 2014, \mnras, 439, 1628,
  \dodoi{10.1093/mnras/stu033}

\bibitem[{{Zu} {et~al.}(2021){Zu}, {Shan}, {Zhang}, {Singh}, {Shao}, {Chen},
  {Yao}, {Golden-Marx}, {Cui}, {Jullo}, {Kneib}, {Zhang}, \& {Yang}}]{Zu2021}
{Zu}, Y., {Shan}, H., {Zhang}, J., {et~al.} 2021, \mnras, 505, 5117,
  \dodoi{10.1093/mnras/stab1712}

\end{thebibliography}
\bibliographystyle{aasjournal}



\end{document}